# Solution of deformed Einstein equations and quantum black holes

**Emre Dil[1], Erdinç Kolay[2]**

[1]Department of Physics, Sinop University, Korucuk, 57000-Sinop, Turkey

[2]Department of Statistics, Sinop University, Korucuk, 57000-Sinop, Turkey

**Abstract**

Recently one and two-parameter deformed Einstein equations have been studied for extremal quantum black holes which have been proposed to obey deformed statistics by Strominger. In this study, we give a deeper insight to the deformed Einstein equations and consider the solutions of these equations for the extremal quantum black holes. We then represent the implications of the solutions, such that the deformation parameters lead the charged black holes to have a smaller mass than the usual Reissner-Nordström black holes. This reduction in mass of a usual black hole can be considered as a transition from classical to quantum black hole regime.

E-mail address: emredil@sakarya.edu.tr

## 1. Introduction

Recently one and two-parameter deformed Einstein equations which are thought to describe the gravitational fields of extremal quantum black holes, have been studied in the frame work of entropic gravity proposal [1]. Extremal black holes forms by a process in which mass of a charged black hole decreases due to the Hawking radiation. Mass of the black hole reaches a minimum value proportional to its charge and this value equals to $Q/\sqrt{G}$ (or $QM_{Planck}$) [2,3]. On the other hand a black hole is a structure that a mass or energy should be concentrated at a region in which an object must have a velocity above the speed of light in order to escape from the gravitational field of that mass. Then the radius of that region is the Schwarzschild radius $R_s = 2Gm/c^2$.

On the other hand, quantum mechanically a mass can only be localized into a region, reduced Compton wavelength $\lambda = \hbar/mc$. When a mass is localized into the reduced Compton wavelength, then it automatically contains the Schwarzschild radius. It means that the localizing a mass into the reduced Compton wavelength creates a black hole since it is concentrated in a region whose radius is smaller than the Schwarzschild radius. Therefore the mass concentrated into the reduced Compton wavelength is a quantum black hole.

These extremal black holes with a possible minimum mass are quantum mechanically stable objects and useful for studying the quantum mechanics of black holes [4].

Quantum mechanics of black holes is dealt by the extremal black holes. For large $Q$ the black holes are macroscopic and for small $Q$ the black holes are microscopic so that the quantum gravity is needed. In order to obtain the quantum field theoretical description of black holes, extremal black holes are considered to be as point particles [5].

One of the way of studying quantum mechanics of black holes is the scattering of black holes to investigate whether they are bosons, fermions or something else [4,5]. Understanding the quantum statistics obeyed by the black holes is a good idea for solving the quantum black hole puzzle. The leading studies has shown that the statistical description of quantum black holes obey neither Bose nor Fermi statistics. Instead the quantum black holes obey infinite statistics or more generally deformed statistics, since infinite statistics firstly introduced by Greenberg [6,7] is the special case of deformed Bose and Fermi algebra [8].

Therefore the extremal quantum black holes can be considered as deformed bosons or fermions and the statistics obeyed by the extremal quantum black holes is deformed statistics. Moreover the statistical mechanics of the deformed bosons and fermions have been studied in the literature through recent years [9-15]. For a particular class of quantum black holes, one type of deformed gas model can be accompanied according to the physical specifications of the black hole and deformed gas model. Arbitrarily two different deformed gas models have been devoted to the different family of extremal quantum black holes in two recent studies [1].

$q$-Deformed Bose gas model and $(q,p)$-deformed Fermi gas model has been taken into account as the quantum black holes. Then the $q$-deformed and $(q,p)$-deformed Einstein equations have been obtained as the gravitational field equations for these deformed gas models. To obtain the deformed Einstein equations, Verlinde's entropic gravity approach [16] has been applied to the deformed entropy of the considered gas model. Verlinde connects the entropy of a source mass to gravitational field equations with a statistical description and reformulated the equations by an entropy-area law. Verlinde's statistical description of gravity has also been inspired to more studies on modifications of Einstein equations [17-33].

Here we firstly give a brief summary of one and two-parameter deformed Einstein equations then the solutions of the deformed Einstein equations for charged black holes. Since the solutions of standard Einstein equations for charged black holes are the Reissner-Nordström solutions in classical gravity, the solutions of the deformed Einstein equations for charged black holes can be considered in quantum gravity. Lastly the implications of the solutions are represented. These are that the deformation parameters lead the charged black holes to have a smaller mass than the usual Reissner-Nordström black holes. This reduction in mass of a usual black hole can be considered as a transition from classical to quantum black hole regime.

## 2. Deformed Einstein equations

By using the entropy of the deformed gas models in Verlinde's entropic gravity approach, the deformed Einstein equations are obtained to describe the gravitational fields of these deformed objects. For a $q$-deformed Bose gas model which is identified by a $q$-deformed boson algebra [34],

$$a_2 a_2^* - q^2 a_2^* a_2 = 1 \,, \tag{1}$$

$$a_1 a_1^* - q^2 a_1^* a_1 = 1 + (q^2 - 1) a_2^* a_2 \,, \tag{2}$$

$$a_2 a_1 = q a_1 a_2 \,, \tag{3}$$

$$a_2 a_1^* = q a_1^* a_2 \,. \tag{4}$$

Here $a$ and $a^*$ represents the deformed annihilation and creation operators, respectively. $q$ is also a real deformation parameter with $0 \le q < \infty$. The grand partition function of the $q$-deformed boson model is [34]

$$Z = \prod_k \sum_{m=0}^{\infty} (m+1) e^{-\beta \varepsilon_k \{m\}} z^m \,, \tag{5}$$

where $\beta = 1/kT$ and $k$ is the Boltzmann constant, $z = e^{\beta \mu}$ is the fugacity, $\varepsilon_k$ is the energy of the single-particle state, $m$ is the occupation number of the single-particle state, and $\{m\}$ is the deformed occupation number and given by

$$\{m\} = \frac{1 - q^{2m}}{1 - q^2} \,. \tag{6}$$

The deformed entropy of the model is also given as

$$S = \frac{4\pi V (2m)^{3/2}}{h^3 T} E^{5/2} \left[ \frac{5\sqrt{\pi}}{4} z + \frac{5\sqrt{\pi}}{2} \delta(q) z^2 - \frac{\sqrt{\pi}}{2} z \ln z - 2\sqrt{\pi} \delta(q) z^2 \ln z + \cdots \right], \tag{7}$$

where $E = kT$ is the average energy of single particle, $V$ is the volume enclosed by the deformed bosons, $m$ is the mass of deformed bosons, $T$ is the temperature of the model and $\delta(q) = (1/4).\{[3/(1+q^2)^{3/2}] - (1/\sqrt{2})\}$ [34]. The deformed entropy in (7) is used to obtain the one-parameter deformed or equivalently the $q$-deformed Einstein equations for $q$-deformed bosons.

On the other hand, to obtain the two-parameter deformed Einstein equations it is suitable to introduce the $(q,p)$-deformed Fermi gas model whose quantum algebraic structure is given by the equations;

$$c_i c_j = -\frac{q}{p} c_j c_i, \quad i < j, \tag{8}$$

$$c_i c_j^* = -qp \, c_j^* c_i, \quad i \neq j, \tag{9}$$

$$c_i^2 = 0, \tag{10}$$

$$c_1 c_1^* + p^2 c_1^* c_1 = p^{2\hat{N}}, \tag{11}$$

$$c_i c_i^* + q^2 c_i^* c_i = c_{i+1} c_{i+1}^* + p^2 c_{i+1}^* c_{i+1}, \quad i = 1, 2, ..., d-1, \tag{12}$$

$$q^{2\hat{N}} = c_d c_d^* + q^2 c_d^* c_d, \tag{13}$$

and where $c_i$ and $c_i^*$ are fermion annihilation and creation operators, respectively and the total deformed number operator is

$$\sum_{i=1}^{d} c_i^* c_i = [\hat{N}_1 + \hat{N}_2 + ... + \hat{N}_d] = [\hat{N}]. \tag{14}$$

Eigenvalue spectrum of total number operator is given by the following generalized Fibonacci basic integers

$$[n] = \frac{q^{2n} - p^{2n}}{q^2 - p^2}, \tag{15}$$

where $q$ and $p$ are the real positive independent deformation parameters [35]. The deformed entropy of the model is

$$S = \frac{(2\pi m)^{3/2} V}{h^3 T} E^{5/2} \left[ \frac{5}{2} f_{5/2}(z, q, p) - f_{3/2}(z, q, p) \ln z \right], \tag{16}$$

where

$$f_n(z,q,p) = \frac{1}{\left| \ln (q^2 / p^2) \right|} \left[ \sum_{l=1}^{\infty} (-1)^{l-1} \frac{(q^2 z)^l}{l^{n+1}} - \sum_{l=1}^{\infty} (-1)^{l-1} \frac{(p^2 z)^l}{l^{n+1}} \right]. \qquad (17)$$

This deformed entropy in (16) is also used to obtain the two-parameter deformed or equivalently the (q,p)-deformed Einstein equations for (q,p)-deformed fermions.

In order to construct the deformed Einstein equations from the entropies in (7) and (16), Verlinde's proposal is applied to the deformed gas models. The fundamental notion needed to derive the gravity is information in the Verlinde's proposal. It is formally the amount of information associated with the matter and its location, measured in terms of entropy. When matter is displaced in space due to a reason, the result is a change in the entropy and this change causes a reaction force. This force is the gravity being an entropic force as an inertial reaction against the force causing the increase of the entropy [16].

The source of gravity is energy or matter and it is distributed evenly over the degrees of freedom in spacetime. The existence of energy or matter in spacetime causes a temperature in the spacetime. The product of the change of entropy during the displacement of source and the temperature is in fact the work and this work is originally led by the force which is known to be gravity [16].

By using the Verlinde's idea, one and two-parameter deformed Einstein equations are recently derived from the deformed entropies (7) and (16) of the *q*-deformed Bose gas model and (*q,p*)-deformed Fermi gas model, respectively [1]. Eventually the *q*-deformed Einstein equations is given, as [1]

$$\frac{10\pi V (2mE)^{3/2}}{h^3} g(z,q) \left( R_{\mu\nu} - \frac{1}{2} g_{\mu\nu} R \right) = 8\pi G T_{\mu\nu}, \qquad (18)$$

where

$$g(z,q) = \left[ \frac{5\sqrt{\pi}}{4} z + \frac{5\sqrt{\pi}}{2} \delta(q) z^2 - \frac{\sqrt{\pi}}{2} z \ln z - 2\sqrt{\pi} \delta(q) z^2 \ln z + \cdots \right]. \qquad (19)$$

Then the (*q,p*)-deformed Einstein equations is similarly given, as

$$\frac{5 V (2\pi m E)^{3/2}}{2 h^3} F(z,q,p) \left( R_{\mu\nu} - \frac{1}{2} g_{\mu\nu} R \right) = 8\pi G T_{\mu\nu}, \qquad (20)$$

where

$$F(z, q, p) = \frac{5}{2} f_{5/2}(z, q, p) - f_{3/2}(z, q, p) \ln z . \qquad (21)$$

The equations in (18) and (20) are one and two-parameter deformed Einstein equations, respectively, and they are assumed to describe the gravitational fields generated by the extremal quantum black holes which obey the statistics of deformed particles in accordance with the Strominger's proposal.

In the next section, we solve one and two-parameter deformed Einstein equations for a charged extremal black hole, and investigate the implications of the solutions.

## 3. Solution of deformed Einstein equations

Since the underlying statistics of the extremal quantum black holes is known to be the deformed statistics, we admit the particles forming deformed gas models to be the quantum black holes and the corresponding deformed Einstein equations for these deformed particles are assumed to describe the gravitational fields of the quantum black holes of these deformed particles.

We know that the extremal quantum black holes should be charged, because the mass of them should decrease to the minimum value proportional to the charge. The classical charged black holes are treated by the standard Einstein equations and the classical solution of the standard Einstein equations for the charged black holes are known as the Reissner-Nordström solutions. Here we obtain the quantum analogs of the solutions of the Einstein equations for these classical charged black holes.

Deformed version of the Einstein field equations is assumed to describe the geometry of the spacetime surrounding a charged spherical quantum black hole. Therefore we need to solve the deformed Einstein-Maxwell equations for the charged quantum black holes. Because of the spherical symmetry generic form for the metric in 4-dimension is [36]

$$ds^2 = -e^{2\alpha(r,t)}dt^2 + e^{2\beta(r,t)}dr^2 + r^2 d\theta^2 + r^2 \sin^2 \theta \, d\phi^2 . \qquad (22)$$

The deformed Einstein equations for the charged spherical quantum black hole is

$$\Psi^{q,p}\left(R_{\mu\nu} - \frac{1}{2}g_{\mu\nu}R\right) = 8\pi\, G T_{\mu\nu}, \tag{23}$$

where

$$\Psi^{q,p} = \begin{cases} \dfrac{10\pi V(2mE)^{3/2}}{h^3}\, g\,(z,q) & \textit{for } q-\textit{deformed Einstein equations} \\[3mm] \dfrac{5V(2\pi\, mE)^{3/2}}{2h^3}\, F(z,q,p) & \textit{for } (q,p)-\textit{deformed Einstein equations} \end{cases}. \tag{24}$$

The energy-momentum tensor $T_{\mu\nu}$ here is one for electromagnetism in this problem and

$$T_{\mu\nu} = F_{\mu\rho}F_\nu^{\ \rho} - \frac{1}{4}g_{\mu\nu}F_{\rho\sigma}F^{\rho\sigma}, \tag{25}$$

where $F_{\mu\nu}$ is the electromagnetic field strength tensor [36]. Also trace of $T_{\mu\nu}$ for $F_{\mu\nu}$ is

$$T = g^{\mu\nu}T_{\mu\nu} = g^{\mu\nu}F_{\mu\rho}F_\nu^{\ \rho} - \frac{1}{4}g^{\mu\nu}g_{\mu\nu}F_{\rho\sigma}F^{\rho\sigma} = 0, \tag{26}$$

since $g^{\mu\nu}g_{\mu\nu} = 4$ in 4-dimensions. Taking the trace of (23) gives $\Psi^{q,p}R = -8\pi\, GT$, then by using this and (26) in (23) gives

$$\Psi^{q,p}R_{\mu\nu} = 8\pi\, G T_{\mu\nu}. \tag{27}$$

Since there is spherical symmetry and only electric charge for our quantum black hole, the electromagnetic field strength tensor has no magnetic field components and the only non-zero components of electric field is radial component which should be independent of $\theta$ and $\phi$. Then the radial electric field component is in the form of

$$E_r = F_{tr} = -F_{rt} = f(r,t). \tag{28}$$

The non-zero components of the Ricci tensor for the metric (22) are given as [36]

$$R_{tt} = \partial_t^2\beta + (\partial_t\beta)^2 - (\partial_t\alpha)(\partial_t\beta) + e^{2(\alpha-\beta)}\left[\partial_r^2\alpha + (\partial_r\alpha)^2 - (\partial_r\alpha)(\partial_r\beta) + \frac{2}{r}\partial_r\alpha\right], \tag{29}$$

$$R_{rr} = -\partial_r^2\alpha - (\partial_r\alpha)^2 + (\partial_r\alpha)(\partial_r\beta) + \frac{2}{r}\partial_r\beta + e^{-2(\alpha-\beta)}\left[\partial_t^2\beta + (\partial_t\beta)^2 - (\partial_t\alpha)(\partial_t\beta)\right], \tag{30}$$

$$R_{tr} = \frac{2}{r}\partial_t\beta, \tag{31}$$

$$R_{\theta\theta} = e^{-2\beta}\big[r(\partial_r\beta - \partial_r\alpha) - 1\big] + 1\,, \tag{32}$$

$$R_{\phi\phi} = R_{\theta\theta}\sin^2\theta\,. \tag{33}$$

Also the corresponding non-zero components of the energy-momentum tensor which is obtained by (25) and (28), are given as [36]

$$T_{tt} = \frac{f(r,t)^2}{2}e^{-2\beta}\,, \tag{34}$$

$$T_{rr} = -\frac{f(r,t)^2}{2}e^{-2\alpha}\,, \tag{35}$$

$$T_{tr} = 0\,, \tag{36}$$

$$T_{\theta\theta} = \frac{r^2 f(r,t)^2}{2}e^{-2(\alpha+\beta)}\,, \tag{37}$$

$$T_{\phi\phi} = T_{\theta\theta}\sin^2\theta\,. \tag{38}$$

By using the two sets of equations above in (29-33) and (34-38), it is also obtained that $\beta(r,t) = \beta(r)$ and

$$\alpha(r,t) = \alpha(r) = -\beta(r)\,. \tag{39}$$

Now the solution of the Maxwell equations $g^{\mu\nu}\nabla_\mu F_{\nu\sigma} = 0$ and $\nabla_{[\mu}F_{\nu\rho]} = 0$ are needed to determine the components of the electromagnetic field strength tensor, $f(r,t)$, in (28). Solving the Maxwell equations for (28) gives

$$f(r,t) = f(r) = \frac{Q}{\sqrt{4\pi}}\frac{1}{r^2}\,. \tag{40}$$

Final step to obtain the solution of the deformed Einstein equations for a charge $Q$ quantum black hole is to find the remaining unknown variable $\alpha(r)$ appearing in the metric (22) for the spacetime which is curved by the charged quantum black hole. To this end, one equation is enough to determine the unknown variable. It can be the $\theta\theta$ component of the deformed Einstein equations (23),

$$\Psi^{q,p}R_{\theta\theta} = 8\pi G T_{\theta\theta}\,, \tag{41}$$

and

$$\partial_r(re^{2\alpha}) = 1 - \frac{1}{\Psi^{q,p}}\frac{GQ^2}{r^2}. \tag{42}$$

The solution is found to be

$$e^{2\alpha} = 1 - \frac{R_S}{r} + \frac{1}{\Psi^{q,p}}\frac{GQ^2}{r^2}, \tag{43}$$

where $R_S$ is the integration constant and known to be the Schwarzschild radius $R_S = 2Gm$. Rewriting the metric (22) with (43) gives

$$ds^2 = \Delta dt^2 + \Delta^{-1}dr^2 + r^2 d\theta^2 + r^2\sin^2\theta\, d\phi^2, \tag{44}$$

where

$$\Delta = 1 - \frac{2Gm}{r} + \frac{1}{\Psi^{q,p}}\frac{GQ^2}{r^2}. \tag{45}$$

The singularities and the event horizons for these black holes are determined by the function $\Delta$ and the radius $r$. There is a true curvature singularity at $r = 0$, since the metric goes to infinity for this value. The coordinate singularity also occurs at $\Delta = 0$ and the conditions giving this singularity occur from the solution of $\Delta = 0$, such as

$$r_{\pm} = Gm \pm \sqrt{G^2 m^2 - \frac{GQ^2}{\Psi^{q,p}}}. \tag{46}$$

This implies that for some suitable cases we can have the event horizons $r_+$ and $r_-$ which determine the place of the coordinate singularity in the spacetime. (46) constitutes three cases of solutions such that $Gm^2 < Q^2/\Psi^{q,p}$, $Gm^2 > Q^2/\Psi^{q,p}$ and $Gm^2 = Q^2/\Psi^{q,p}$.

First case $Gm^2 < Q^2/\Psi^{q,p}$ is unphysical since this solution states that the total energy of the black hole is less than the energy of the electromagnetic contribution. Also this condition makes $\Delta$ different from zero, which makes the first case invalid.

Second case $Gm^2 > Q^2/\Psi^{q,p}$ implies a physical situation since the energy of electromagnetic field is less than the total energy. Two event horizons $r_+$ and $r_-$ also make $\Delta = 0$.

Finally the third case $Gm^2 = Q^2 / \Psi^{q,p}$ gives the extremal charged black hole solution, since the mass of the black hole decreases to the minimum value from the second case $Gm^2 > Q^2 / \Psi^{q,p}$. This minimum mass solution for extremal black holes remains stationary for all times. Also this case makes $\Delta = 0$ at a single radius $r_{\pm} = Gm$ and this states a single event horizon. This deformed case solution $Gm^2 = Q^2 / \Psi^{q,p}$ is the analogue of classical Reissner-Nordström solution $m = Q / \sqrt{G}$ which is often examined in the studies of quantum gravity. In the second case, the mass of the black hole is allowed to be in very large classical scales due to the ability of getting bigger values than the charge, implied in the inequality $Gm^2 > Q^2 / \Psi^{q,p}$. Whereas the mass of the deformed black hole is allowed to decrease very small values which could fall into the quantum regime, because the decrease of the mass is governed by a very small term $1 / \Psi^{q,p}$ being order of $h^{6/7}$ in the right hand side of the third case equation $Gm^2 = Q^2 / \Psi^{q,p}$.

In our deformed case, this decrease in mass of black hole which is controlled by the term $1 / \Psi^{q,p}$ is different from the classical Reissner-Nordström solution. We now discuss the effects of this extra term on mass reduction.

We investigate the reduction of the mass with respect to the classical Reissner-Nordström case, for the $q$-deformed and ($q,p$)-deformed Einstein cases. From (24), we have two $1 / \Psi^{q,p}$ values for the mass of the extremal quantum black hole in the third case of $Gm^2 = Q^2 / \Psi^{q,p}$, such as

$$m^q = \left( \frac{h^3}{10\pi V (2E)^{3/2}} \frac{1}{g(z,q)} \right)^{\frac{2}{7}} \left( \frac{Q^2}{G} \right)^{\frac{2}{7}}, \tag{47}$$

$$m^{(q,p)} = \left( \frac{2h^3}{5V(2\pi E)^{3/2}} \frac{1}{F(z,q,p)} \right)^{\frac{2}{7}} \left( \frac{Q^2}{G} \right)^{\frac{2}{7}}. \tag{48}$$

While the minimum mass of an extremal quantum black hole for the $q$-deformed case is (47), it is (48) for the ($q,p$)-deformed case. However, the minimum mass of a classical

Reissner-Nordström black hole is given as $m = Q/\sqrt{G}$. When we compare the minimum masses of deformed quantum case and the classical Reissner-Nordström cases, we obtain

$$m^q = \left( \frac{1}{10\pi V(2E)^{3/2} Q^{3/2}} \right)^{\frac{2}{7}} \left( \frac{h^3 G^{3/4}}{g(z,q)} \right)^{\frac{2}{7}} m \,, \tag{49}$$

$$m^{(q,p)} = \left( \frac{2}{5V(2\pi E)^{3/2} Q^{3/2}} \right)^{\frac{2}{7}} \left( \frac{h^3 G^{3/4}}{F(z,q,p)} \right)^{\frac{2}{7}} m \,. \tag{50}$$

These equations in (49) and (50) imply that the mass of the charged extremal black hole in the deformed quantum case can decrease to a smaller value than that of the classical Reissner-Nordström case. To understand the decrease in the mass, we examine the behaviors of the factors $(h^3 G^{3/4}/g(z,q))^{2/7}$ and $(h^3 G^{3/4}/F(z,q,p))^{2/7}$ in front of the classical mass of the charged black hole in (49) and (50), respectively. Therefore, it is represented that the behavior of $(h^3 G^{3/4}/g(z,q))^{2/7}$ with respect to $z$ and $q$ in Fig. 1 and Fig. 2, for $q < 1$ and $1 < q$, respectively. We also represent the behavior of $(h^3 G^{3/4}/F(z,q,p))^{2/7}$ with respect to $z$, $q$ and $p$ in Fig. 3 and Fig. 4, for $q,p < 1$ and $1 < q, p$, respectively.

## 4. Conclusions

Recently, $q$-deformed and $(q,p)$-deformed Einstein equations have been proposed for the investigations of charged extremal quantum black holes, in the framework of entropic gravity approach [1]. In this study, we give a review and deeper meaning to the deformed Einstein equations, which is based on the Strominger's idea, such that the quantum black holes obey the deformed statistics. We then consider the solutions of these equations for the charged extremal quantum black holes. We analyze the obtained solutions for $q$-deformed and $(q,p)$-deformed cases, separately.

We represent the true and coordinate singularities from the solutions for quantum black holes. Also the event horizons for these singularities are mentioned briefly. We also investigate the possible decrease in mass via Hawking radiation to a minimum value which is determined by the charge of quantum black hole. The difference between the decrease in classical black holes and quantum black holes is obvious from the equations (47), (48) and $m = Q / \sqrt{G}$. According to this difference, the reduced quantum and classical mass of the extremal black holes are represented in (49) and (50).

We illustrate the decreases in quantum masses $m^q$ and $m^{q,p}$ in Figs. 1-4 with respect to the classical mass $m$. According to the Figs. 1-2, the mass of the quantum black hole $m^q$ in (49) is at least $10^{-30}$ times smaller than the classical black hole mass $m$, in the $q$-deformed case. After considering the inverse of the volume, charge and energy factors, mass $m^q$ gets smaller than $10^{-30} m$. We again see a similar situation for the mass of the quantum black hole $m^{q,p}$ in (50) from Figs. 3-4. $m^{q,p}$ is at least $10^{-30}$ times smaller than the classical black hole mass $m$, in the $(q,p)$-deformed case. Considering the inverse of the volume, charge and energy factors, mass $m^{q,p}$ similarly gets smaller than $10^{-30} m$ in the $(q,p)$-deformed case.

Since the theoretical possibility of concentrating a mass into its reduced Planck mass gives a radius containing the Schwarzschild radius and the obtained quantum masses of the extremal black holes in (49) and (50) are at least $10^{-30}$ times smaller than the classical masses due to a possible Hawking radiation, the solutions of the deformed Einstein equations imply that the all the propositions and ideas considered here seem consistent with each other. Because there have been used three independent ideas to obtain these equations. Verlinde's proposition is on gravity having an entropic origin, Strominger's is on the type of the underlying statistics obeyed by the quantum black holes, and our idea is to get the gravitational field equations for these quantum black holes from Verlinde's proposition, by considering the quantum black holes as the deformed bosons or fermions due to the Strominger's statement that the statistics obeyed by the quantum black holes is deformed statistics.

**Conflict of Interests**

The authors declare that there is no conflict of interest regarding the publication of this paper.

**Figure Captions**

**Figure 1.** The $q$-deformed mass reduction factor $(h^3 G^{3/4} / g(z,q))^{2/7}$ for $q<1$.

**Figure 2.** The $q$-deformed mass reduction factor $(h^3 G^{3/4} / g(z,q))^{2/7}$ for $q>1$.

**Figure 3.** The $q,p$-deformed mass reduction factor $(h^3 G^{3/4} / F(z,q,p))^{2/7}$ for various values of the deformation parameters $p$ and $q<1$.

**Figure 4.** The $q,p$-deformed mass reduction factor $(h^3 G^{3/4} / F(z,q,p))^{2/7}$ for various values of the deformation parameters $p$ and $q>1$.

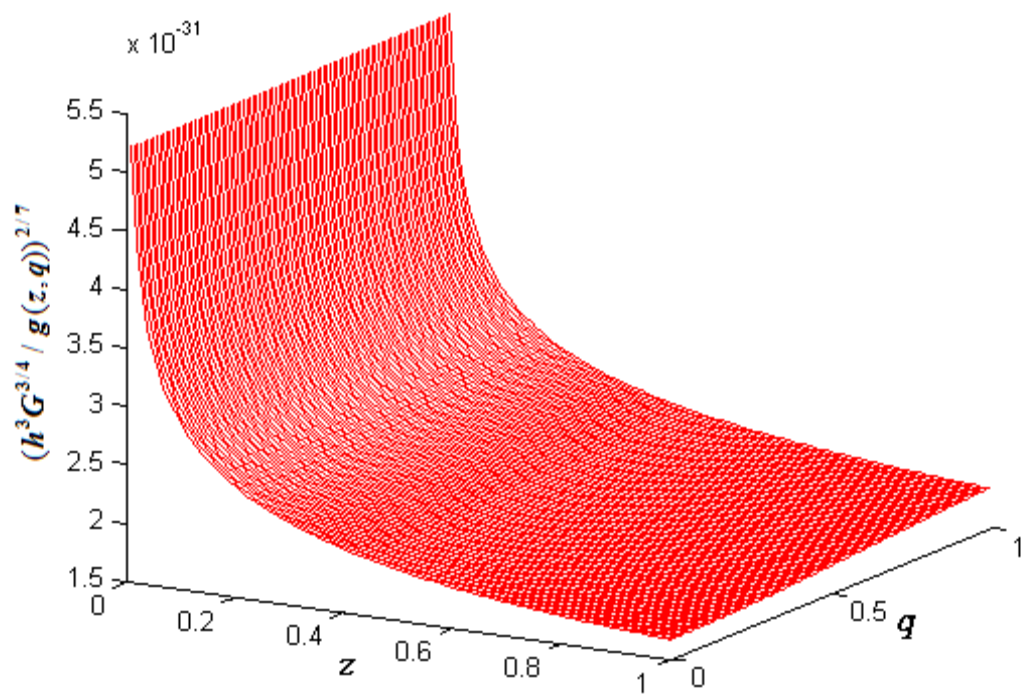

Figure 1.

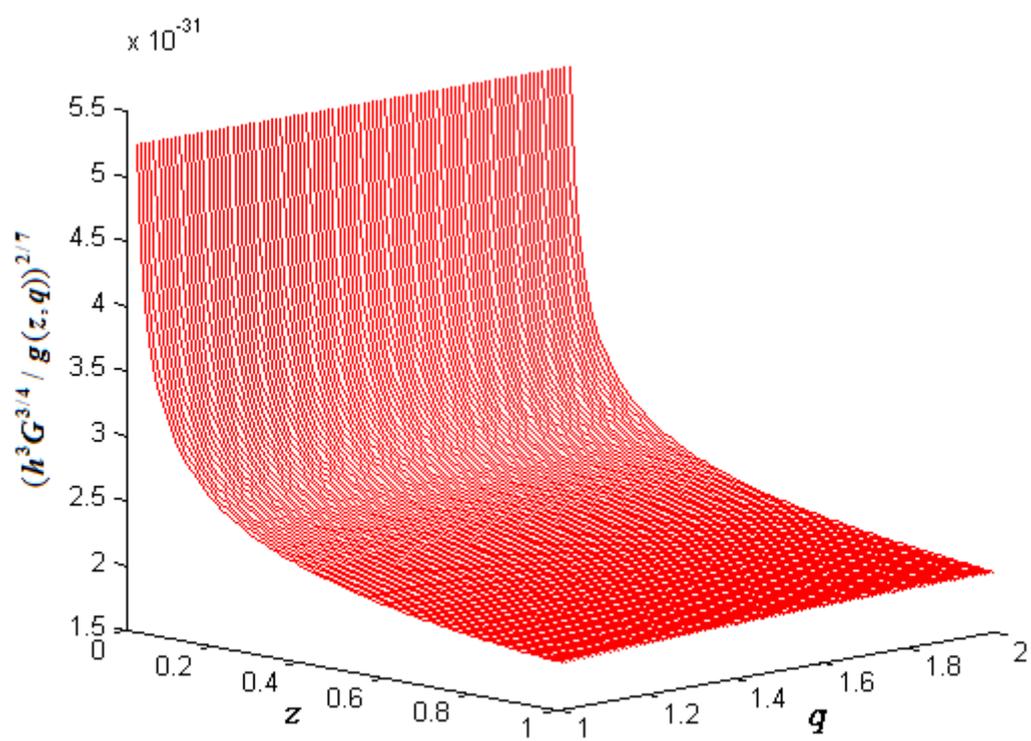

Figure 2.

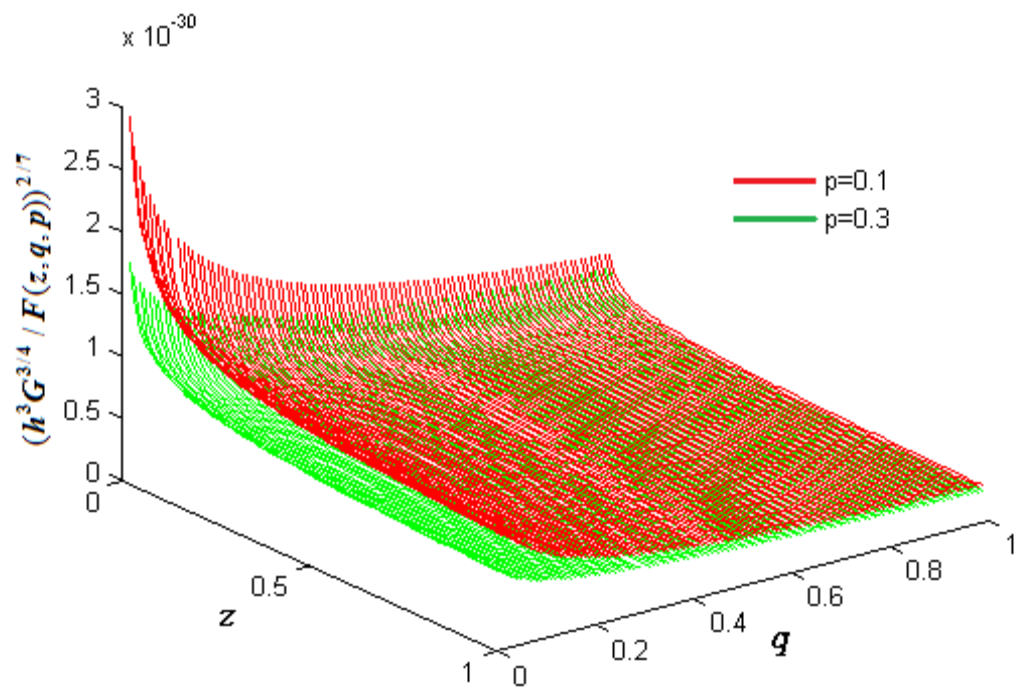

Figure 3.

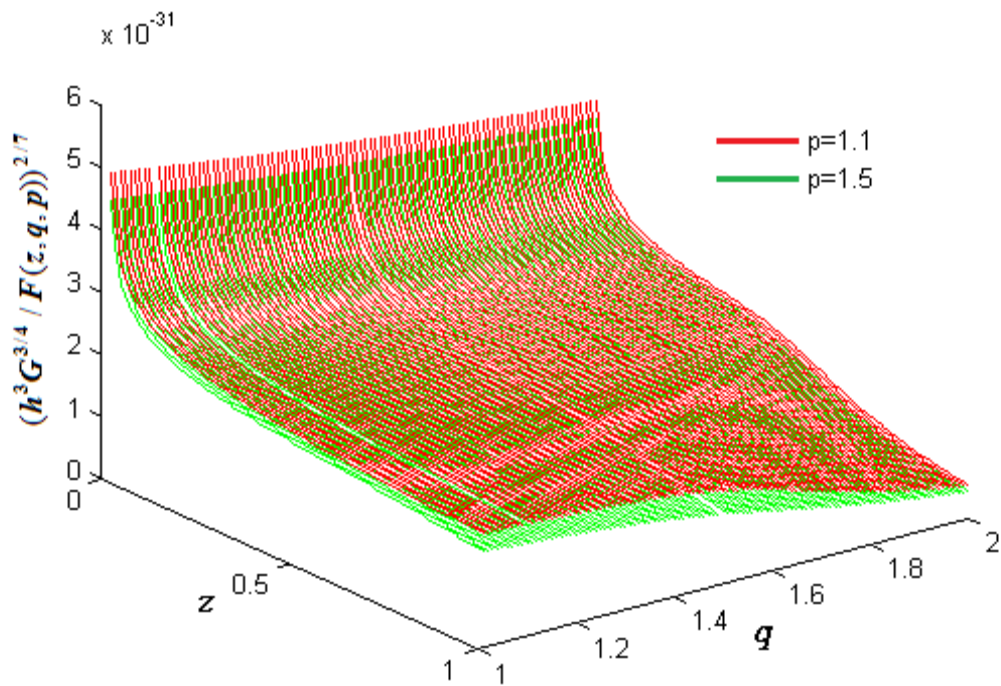

Figure 4.